# Do conformational changes contribute to the surface plasmon resonance signal?

Daniel Dobrovodský[1, 2] & Carmelo Di Primo[3*]

1. Institute of Biophysics, Czech Academy of Sciences, Královopolská 135, 612 65 Brno, Czech Republic

2. National Centre for Biomolecular Research, Faculty of Science, Masaryk University, Kamenice 5, 625 00 Brno, Czech Republic

3. Univ. Bordeaux, CNRS, INSERM, ARNA, UMR 5320, U1212, IECB, F-33000 Bordeaux, France

*To whom correspondence should be addressed. Email: carmelo.diprimo@inserm.fr, Tel: +33 5 40 00 30 46, https://orcid.org/0000-0002-0509-8399.

The authors declare no conflict of interest

The authors confirm contribution to the paper as follows: study conception and design: C. Di Primo; data collection: D. Dobrovodský; analysis and interpretation of results: D. Dobrovodský, C. Di Primo; draft manuscript preparation: D. Dobrovodský; correction and preparation of the submitted version of the manuscript: D. Dobrovodský, C. Di Primo. All authors reviewed the results and approved the final version of the manuscript.

**Abstract**

Surface plasmon resonance (SPR)-based biosensors are widely used instruments for characterizing molecular interactions. In theory the SPR signal depends only on mass changes for interacting molecules of same chemical nature. Whether conformational changes of interacting molecules also contribute to the SPR signal is still a subject of lively debates. Works have been published claiming that conformational changes were detected but all factors contributing to the SPR signal were not carefully considered, in addition to often using no or improper controls. In the present work we used a very well-characterized oligonucleotide, the thrombin-binding DNA aptamer (TBA), which upon binding of potassium ions folds into a two G-tetrad antiparallel G-quadruplex structure. All terms contributing to the maximal expected SPR response, $R_{max}$, in particular the refractive index increment, RII, of both partners and the fraction of immobilized TBA target available, $c_a$, were experimentally assessed. The resulting $R_{max}$ was then compared to the maximal experimental SPR response for potassium ions binding to TBA using appropriate controls. Regardless how the RIIs were measured, by SPR or refractometry, and how much TBA available for interacting with potassium ions was considered, the theoretical and the experimental SPR responses never matched, the former being always lower than the latter. Using a straightforward experimental model system and by thoroughly taking into account all contributing factors we therefore conclude that conformational changes can indeed contribute to the measured SPR signal.

**Abbreviations**

$c_a$: Fraction of active target

CD: Circular dichroism

G4: G-quadruplex

MW: Molecular weight

ODN: Oligodeoxynucleotide

RB: Running buffer

RII: Refractive index increment

RU: Resonance units

S: Stoichiometry

SA: Streptavidin

SCK: Single cycle kinetics

SPR: Surface plasmon resonance

TBA: Thrombin binding aptamer

TEG: Triethylene glycol

## 1. Introduction

In the past decades, surface plasmon resonance (SPR)-based biosensors have been firmly established as "gold" standard instruments for analysing molecular interactions. The technique allows for label-free, real-time qualitative monitoring of molecular binding as well as quantitative evaluation of interaction parameters. For these reasons, SPR has found applications in life sciences where it has been used to study the affinity, specificity, and binding kinetics of protein-protein (Meyerkord and Fu, 2015), protein-nucleic acid (Majka and Speck, 2006; Šípová and Homola, 2013), and nucleic acid-nucleic acid complexes (Palau et al., 2013; Šípová and Homola, 2013). SPR has also been used to investigate enzymatic reactions (Wang et al., 2008), receptor-drug interactions (Ma et al., 2018), as well as for screening for specific small-molecular ligand (Vo et al., 2019) or aptamers (Dausse et al., 2016) and other purposes (Nguyen et al., 2015).

SPR is based on the measurement of refractive index changes in the proximity of the sensing surface as a result of the immobilization of one molecule, the ligand, onto the surface and its subsequent interaction with its partner, the analyte. This generates the SPR response, which on Biacore instruments is measured in resonance units (RU). In most practical applications, these changes were considered solely as a function of mass increment occurring at the solid-liquid interface of the biochip. However, the SPR signals upon immobilization of the ligand and interaction with the injected analyte depend on multiple factors, according to Eq. 1 (Davis and Wilson, 2000):

$$\frac{R_1}{R_{max}} = \frac{MW_1}{MW_2} \times \frac{RII_1}{RII_2} \times S \qquad (1)$$

where $R_1$ is the response after immobilization of the ligand, $R_{max}$ is the maximal response at saturating concentration of the analyte, $MW_1$ and $MW_2$ are the molecular weights of the ligand and analyte, respectively, $RII_1$ and $RII_2$ are the corresponding refractive index increments of the two species, and S refers to the stoichiometry of the interaction, i.e., the number of binding sites of the ligand. With all the terms known, Eq. 1 allows convenient prediction of expected signal response, $R_{max}$, upon analyte binding. However, not all immobilized target is generally available for analyte binding due to its partial degradation, denaturation, steric effects, or dissociation from the surface, which further influences the maximal obtainable signal of the analyte. For that reason, an availability constant, $c_a$, reflecting the fraction of available target must be included in $R_{max}$ prediction. The final expression is then given by Eq. 2:

$$R_{max} = R_1 \times \frac{MW_2}{MW_1} \times \frac{RII_2}{RII_1} \times \frac{1}{S} \times c_a \qquad (2)$$

Since the beginning of the use of SPR in biomolecular analysis, a question has been raised, whether the binding-associated conformational change in a studied system also contributed to the SPR signal. Many works have been published on this topic claiming to detect conformational changes of biomolecules using SPR (Salamon et al., 1994; Sota et al., 1998; Boussaad et al., 2000; Mannen et al., 2001; Zako et al., 2001; Flatmark et al., 2001; Gestwicki et al., 2001; May and Russel, 2002; Hsieh et al., 2002; Geitmann and Danielson, 2004; Wood et al., 2005; Christopeit et al., 2009; Dell'Orco et al., 2010, 2012; Hall et al., 2011; Dejeu et al., 2018, 2021). These claims were usually based on changes in baseline of immobilised proteins upon induction of conformational change by various agents (pH, electric potential, denaturing buffer) (Boussaad et al., 2000; Mannen et al., 2001; May and Russell, 2002; Salamon et al., 1994; Sota et al., 1998; Zako et al., 2001) or on observations of additional SPR signal for interactions of proteins with small molecules or ions, that could not be compensated by the increase of mass alone and thus was attributed to changes in conformation (Christopeit et al., 2009; Dejeu et al., 2021, 2018; Dell'Orco et al., 2012, 2010; Flatmark et al., 2001; Geitmann and Danielson, 2004; Gestwicki et al., 2001; Hsieh et al., 2002). However, these results are controversial since the authors failed to consider all factors contributing to the SPR signal, in addition to often using no or improper experimental controls. Although some studies attempted to calculate the expected signal from an equivalent of Eq. 2 (Dejeu et al., 2021, 2018; Gestwicki et al., 2001; Hsieh et al., 2002; Wood and Lee, 2005), the authors attributed incorrect values to the ratio of refractive index increments or neglected it entirely. Therefore, the uncompensated signal observed in these studies cannot be unequivocally attributed to a change in conformation.

The present work aimed to properly determine whether conformational changes of biomolecules can be detected using SPR and contribute to the measured response. For this purpose, a very well-characterized oligodeoxynucleotide (ODN), the thrombin-binding DNA aptamer (Bock et al., 1992; Macaya et al., 1993; Paborsky et al., 1993; Wang et al., 1993), was chosen that is able, upon interaction with $K^+$ ion to fold into a G-quadruplex (G4), a non-canonical compact structure of DNA (Ma et al., 2020). This G4 aptamer was immobilized on a sensor chip through streptavidin-biotin coupling (Fig. 1a). The calculated maximal expected SPR signal, $R_{max}$, upon potassium binding was predicted from Eq. 2 by careful experimental evaluation of all its terms and using appropriate control ODNs. The theoretical $R_{max}$ was then compared to the maximal experimental SPR signal obtained from the interaction of the ODN with $K^+$ inducing folding of the G4 structure. These $R_{max}$ values did not match. After thoroughly considering all known contributing factors, any discrepancy between these two values could be attributed to the conformational change taking place when the ODN folds from a single strand to a G4. Therefore, we conclude that conformational changes can contribute to the SPR signal measured on Biacore instruments and the most widely used SPR biosensors in academic and private laboratories.

## 2. Materials and methods

### 2.1. Materials

High purity salt-free thrombin-binding aptamer (TBA; 5'-GGTTGGTGTGGTTGG-3'), a control sequence (TBAcont; 5'- AATTGGTGTAATTGG -3') with or without a biotin tag attached to their 5' end via a triethylene glycol (TEG) spacer as well as complementary strands TBAcomp (5'-CCAACCACACCA-3') and TBAcontcomp (5'-CCAATTACACCA-3') were purchased from Eurofins Genomics (Nantes, France). For the refractometry measurements, the TBA sequence was synthesised at a high synthesis scale in-house using an ÄKTA-O-10 synthesizer (Cytiva, Umea, Sweden). Before use, all ODNs were desalted by running them through Sephadex G-25 spin columns (GE Healthcare, Vélizy-Villacoublay, France). Their concentration was determined in Milli-Q water by absorbance measurement using a NanoDrop One spectrophotometer (Thermo Scientific, Illkirch, France). The stock solutions were stored at −20 °C.

2.2. Surface plasmon resonance experiments

All SPR experiments were performed on a Biacore T200 instrument (Cytiva, Uppsala, Sweden) at 25 °C.

The interaction experiments with KCl were performed on SAHC200M sensor chips (Xantec Bioanalytics, Düsseldorf, Germany) with a layer of streptavidin immobilized on a carboxy-methyl-dextran-modified gold surface. The surface of a new sensor chip was first cleaned by three consecutive 30 s injections of a solution of 50 mM NaOH and 1 M NaCl prepared in Milli-Q water to remove any non-covalently associated streptavidin. The biotinylated ODNs were immobilized on flow cells 2 and 4 by injection of 50 nM solutions in HBS-EP+ buffer (10 mM HEPES, pH 7.4, 150 mM NaCl, 3 mM EDTA, 0.05% (v/v) Surfactant P20) until the desired amount of immobilized ODN was reached (ca. 1000 RU). Channels 1 and 3 were left empty and used as a reference. As a running buffer (RB) for the binding experiments, 100 mM HEPES was used with pH adjusted to 7.2 with LiOH. To avoid possible contamination of the buffer solution by $K^+$ from the pH electrode a first solution was adjusted to pH 7.2 using the pH-meter to know exactly how much LiOH was required. Then another solution was prepared without using the pH-meter. Solutions of KCl with increasing concentrations in the range from 0.9375 mM to 30 mM in RB were then injected for 30 s at 25 µl min$^{-1}$ followed by 60 s dissociation. In between injections, the surface was cleaned with RB for 60 s.

The amount of active target available for binding was determined by immobilizing the biotinylated target ODNs on the SAHC200M sensor chip and subsequent injection of the complementary strand in RB at 25 µl min$^{-1}$ using the single cycle kinetics method (SCK) (Karlsson et al., 2006; Palau and Di Primo, 2012), with 60 s and 400 s contact and dissociation times, respectively. For the regeneration step, 20 mM LiOH in milliQ water was applied for 60 s. All experiments were performed with duplicate injections and the obtained sensorgrams were double referenced to the blank buffer and to the response of the empty reference channel.

Refractive index determination by SPR was performed by injection of the measured solution over a cleaned, unmodified surface of the SAHC200M sensor chip or the plain gold surface of AU sensor chip (Xantec Bioanalytics, Düsseldorf, Germany) for 30 s at flow rate 25 µl min$^{-1}$. The refractive index values were determined from the sensorgrams by reading of the response value 5 s before the end of the injection.

2.3. Refractometry

Measurements of refractive index data were carried out at 589 nm using Abbe refractometer DR-A1-Plus (ATAGO, USA) at a controlled temperature of 25 °C. Refractive index of the pure solvent (100 mM HEPES pH 7.2) was subtracted from the value measured for each solution to obtain the change in refractive index change (Δn) corresponding to the dissolved species.

2.4. Circular dichroism spectroscopy

Circular dichroism (CD) spectroscopy was used to monitor titration of ODNs with K$^+$ ions. CD spectra were recorded on Jasco J-1500 CD spectrophotometer (Jasco, Lisses, France) in a Peltier thermostated holder at 25 °C using 1.0 cm quartz cuvettes in the wavelength range 220 – 330 nm with 50 nm min$^{-1}$ scan speed and 2 nm bandwidth. 5 µM solution of ODN in 100 mM HEPES pH 7.2 was titrated with additions of 100 mM KCl in 100 mM HEPES pH 7.2 with 5 µM of the appropriate ODN to avoid dilution of the ODN in the cuvette. The titration continued until final concentration of 30 mM KCl was reached. The solution was left to equilibrate after each addition for 5 min prior to the measurement. Each spectrum was obtained as an average of three successive accumulations.

2.5. Thermal melting experiments

Thermal denaturation and renaturation of TBA and control ODN was carried out using a UV-mc$^2$ spectrophotometer (SAFAS, Monaco) with Peltier temperature-controlled cuvette holder. The TBA samples prepared at 5 µM strand concentration in 100 mM HEPES pH 7.2 with or without 30 mM KCl were heated to 90 °C for 90 s and subsequently left at room temperature for 10 min to facilitate the formation of G4 before being placed to 1 cm quartz cuvettes and capped by paraffin oil for the measurement. The melting and renaturation curves were recorded from 5 °C to 90 °C at 0.3 °C min$^{-1}$. The structural transition was monitored by measuring the absorbance at 295 nm. The recorded melting curves were referenced to the blank buffer solution as well as to TBA absorbance at control wavelength 320 nm. Melting temperatures ($T_m$) were determined as the inflection point of the melting curve from its first derivative.

## 3. Results

3.1. CD-monitored titration of ODNs with $K^+$

In a previous work (Di Primo and Lebars, 2007) it was shown that TBA immobilized on a sensor chip folded as a G4 in the presence of potassium as it was able to recognize the target of the *in vitro* selection (Bock et al., 1992), thrombin. To confirm again the suitability of TBA as a conformational change model, titration of TBA by $K^+$ was performed using CD spectroscopy, at the same buffer conditions and KCl concentration range (0 – 30 mM) used in the SPR experiments. The same experiment was performed with TBA carrying biotin-TEG tail which was used for immobilisation of ODNs on a streptavidin coated sensor chip in the SPR experiments, to assess the effect of this covalent tag on the formation of G4. For control experiments, similar $K^+$ titrations were performed with a TBAcont (biotinylated and non-biotinylated) sequence. TBAcont was designed to not form G4, due to substitution of two guanines of the TBA sequence with adenines.

A relationship between various G4 conformations and position of bands in CD spectra has been well-established empirically (Karsisiotis et al., 2011) but also rationalised theoretically (Masiero et al., 2010), making it an excellent method for the analysis of G4 structures. TBA displayed a CD spectrum as previously described for this sequence characteristic for antiparallel G4 (Nagatoishi et al., 2007), with a negative band around 270 nm and two positive bands at 245 and 293 nm (Fig. 1). Gradual increase of the latter positive peak for the aptamer without (Fig. 1b) and with biotin-TEG (Fig. 1c) with increasing KCl concentration was used to follow the course of the titration reflecting G4 formation (Fig. 1d).

Significant increase of this signal was observed for TBA already after the first addition of KCl. The signal reached saturation at around 10 mM KCl (Fig. 1b). In the case of TBA with the biotin-TEG tail, the increase was more gradual, with higher concentration of KCl needed for the G4 to be fully folded. Nevertheless, saturation was reached by the end of the titration. The maximum signal intensity reached at saturation was also lower than for the unmodified TBA (Fig. 1c), possibly indicating lower stability of the modified ODN (Nagatoishi et al., 2007). Additionally, Fig. 1d shows notably larger standard deviation from three experiments for biotin-TEG-TBA, likely as a result of the higher instability of this sample.

The titration data were fitted to a theoretical model to obtain the apparent dissociation equilibrium constant, $K_D$, for $K^+$ binding to TBA. Since the concentration of the latter was well below the used concentration range of KCl and the expected $K_D$, Eq. 3 could be used (Jarmoskaite et al., 2020):

$$R_{eq} = \frac{[KCl] \times R_{max}}{K_D + [KCl]} \qquad (3)$$

where $R_{eq}$ is the obtained response at the equilibrium and $R_{max}$ is the hypothetical maximal response at infinite KCl concentration (Fig. 1d). For TBA carrying the biotin-TEG tail the $K_D$ determined from the fit was 1.6 ± 0.3 mM compared to $K_D$ of 0.20 ± 0.02 mM for the unmodified TBA, indicating partial loss of affinity for $K^+$ ions as a result of the covalent modification.

CD spectra of TBAcont without and with biotin-TEG (Supplementary material, Fig. S1) showed one positive and one negative bands of low intensity, indicating that this ODN is not completely unstructured. However, no conformational transition was observed, as these bands exhibited no response to increasing concentration of KCl. Thus, this sequence was confirmed to be suitable as a non-G4-forming control for TBA.

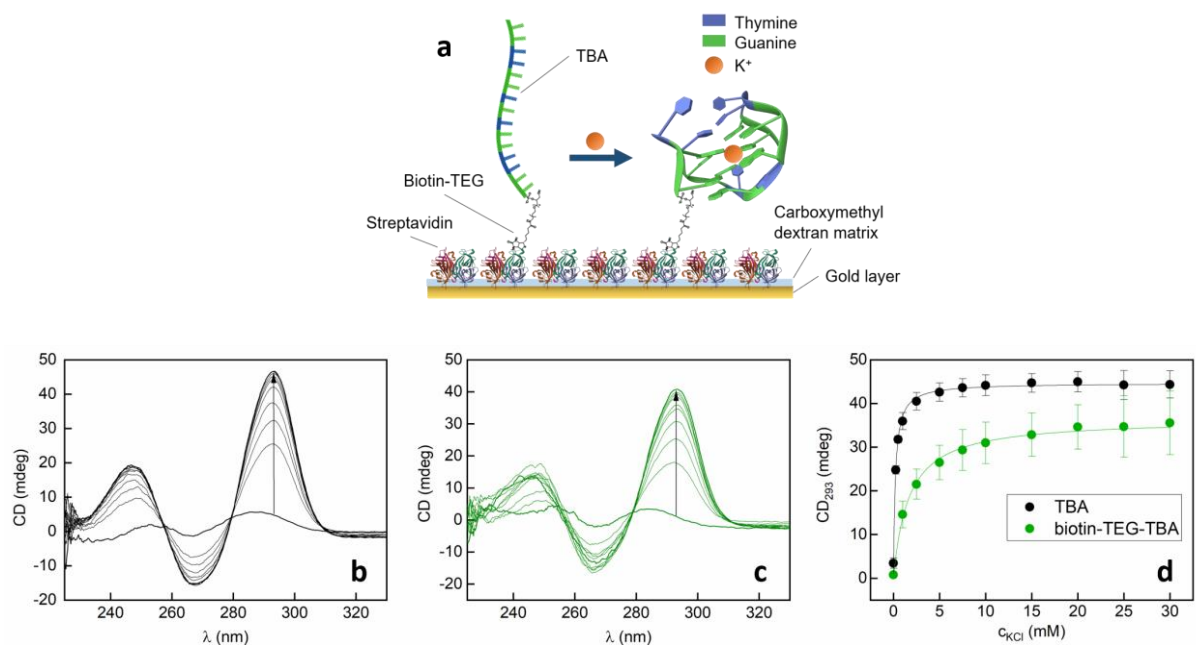

**Figure 1.** (a) Schematic diagram of the experimental design, TBA G4 structure was made from https://www.rcsb.org/structure/148D, 10.2210/pdb148D/pdb (Schultze et al., 1994); CD spectra of the titration of 5 μM TBA (b) and biotin-TEG-TBA (c) with KCl in 100 mM HEPES buffer pH 7.2; arrows indicate increasing concentrations of KCl. (d) Fitted titration curves evaluated at 293 nm; error bars indicate standard deviation from three independent experiments.

3.2. UV absorbance-monitored thermal melting of TBA

Thermal denaturation/renaturation experiments of TBA monitored by UV-spectroscopy were carried out in order to analyse its stability under the conditions used for the SPR experiments. TBA with and without the biotin-TEG tail, which is going to be used to immobilize it on a streptavidin sensor chip, produced melting profiles in the presence of 30 mM KCl with no hysteresis upon renaturation as expected for a two-state folding reaction (Supplementary material, Fig. S2). No transition was observed in the absence of $K^+$ ions, illustrating their crucial role in G4 formation in TBA. From the minimum of first derivatives of the melting curves, $T_m$ values were determined as 45.6 ± 1.3 °C for TBA (Supplementary material, Fig. S2a) and 38.4 ± 0.8 °C for biotin-TEG-TBA (Supplementary material, Fig. S2b), indicating destabilisation of the G4 by addition of the biotin-TEG tail. Even though this method of $T_m$ determination is known to not be very precise (Mergny and Lacroix, 2009), it is adequate enough for the purpose of this study. Furthermore, the value determined for TBA agrees with the values reported in the literature (Mergny et al., 1998; Saccà et al., 2005).

The fraction of folded G4 strands in the presence of KCl was analysed as described by Mergny et al. (Mergny and Lacroix, 2009). At 25 °C, i.e. at the temperature of the SPR experiments, it was estimated to be 0.99 and 0.95 for TBA and biotin-TEG-TBA, respectively, reflecting the lower stability of the latter species. In spite of its lower stability, essentially all biotin-TEG-TBA molecules are folded under the studied conditions. Without $K^+$, at 25 °C, the biotin-TEG TBA is clearly unfolded (Supplementary material, Fig. S2b). These results validate the use of TBA as a very good model to investigate to what extent conformational changes contribute to the SPR signal.

### 3.3. Determination of refractive index increments

Binding of the free analyte to the immobilized interaction partner (ligand) during an SPR experiment causes change of the refractive index in the proximity of the surface and consequently also change of the resonance angle. Therefore, the refractive index increment, RII, of a given specie, defined as $\delta n/\delta c$, i.e. change of refractive index (n) with the concentration (c) of the species, is a crucial parameter to determine the expected maximal response upon binding of the analyte at saturating concentration. This is especially true in systems where the two interaction partners are fairly chemically different, such as in the system presently studied, and the ratio of their RIIs in Eq. 2 may deviate significantly from 1 (Davis and Wilson, 2000; Tumolo et al., 2004).

To determine the RIIs of KCl and TBA, refractometry measurements of two independently prepared sets of samples of increasing concentration of each interaction partner were performed. Additionally, the RII value can be affected by the ionic conditions of the base buffer. To see whether the increasing concentration of $K^+$ ions in SPR experiment would have a significant effect on the RII value of TBA, the TBA solutions were prepared both in the presence or absence of 30 mM KCl. RIIs were determined as slopes of concentration dependence of refractive index (Fig. 2) from two independent sets of measurements (Supplementary material, Table S1). It is clear that RIIs of KCl and TBA differ significantly, and thus in Eq. 2 would produce a ratio vastly distant from 1, highlighting the importance of checking this parameter. The results also show that the presence of 30 mM KCl in the solution has a non-negligible effect on the RII of TBA (Fig. 2a), lowering its RII from 0.268 to 0.222.

In order to check for the consistency of the measured RII values, the same samples used for refractometry were also measured using SPR, taking advantage of the T200 instrument, which is actually also a refractometer. The results are shown in Fig. 2b (corresponding sensorgrams shown in the Supplementary material, Fig. S3b, c, d). In the absence of specific binding, the SPR response reflects only the change of refractive index (Δn) at the wavelength used by the instrument (760 nm) near the surface, resulting in sensorgrams with square profile (Supplementary material, Fig. S3). On the Biacore instrument, a response of 1 RU corresponds to a Δn of $1\times10^{-6}$. Samples were injected over the streptavidin coated sensor chip surface used for the SPR binding experiments. Injection of KCl samples with increasing concentration (Supplementary material, Fig. S3b) resulted in RII value of 0.117, in sufficiently good agreement with the refractometry value, 0.122 (Supplementary material, Table S1). The average RII of KCl, 0.120, was further used for $R_{max}$ calculation. On the other hand, the RII values determined by SPR for TBA on the surface of a sensor chip covered with the streptavidin-carrying hydrogel matrix were lower compared to those obtained by refractometry by about 22 %. The presence of 30 mM KCl in the samples of TBA introduces slight curvature in the steady-state portion of the recorded sensorgams (Supplementary material, Fig. S3d) in contrast to the flat sensorgrams of TBA in clean buffer (Fig. S3c). This might suggest the existence of some kind of interaction of TBA with the immobilized layer, facilitated by the $K^+$ ions. Nonetheless, the difference in obtained RII value arising from the presence of 30 mM KCl is exactly the same as determined by refractometry (0.046). Clearly, determination of the appropriate RII value for TBA is not straightforward and is further discussed below. Ultimately, for the purpose of this study, all measured values have to be considered.

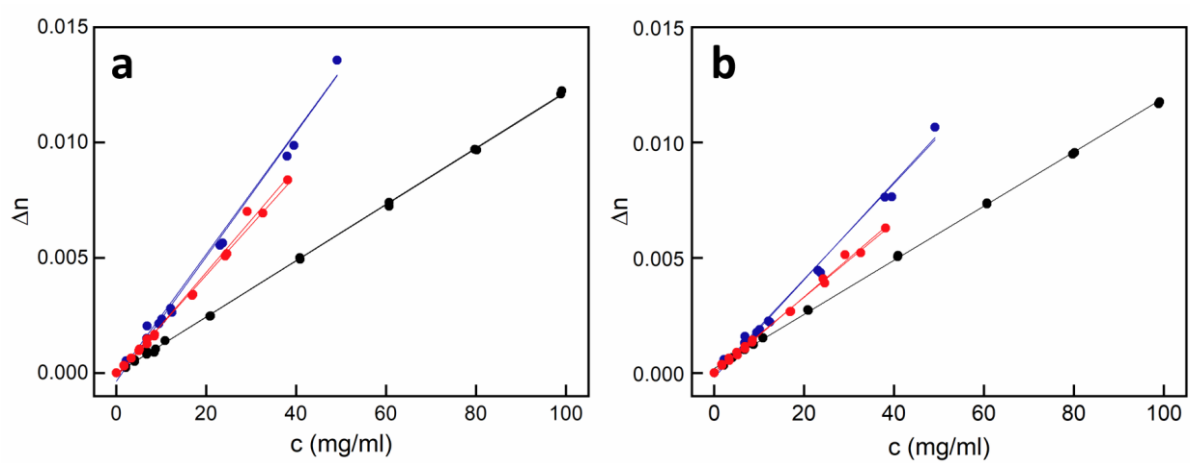

**Figure 2.** Dependencies of refractive index change on concentration of KCl (black) and TBA in the presence (red) or absence (blue) of 30 mM KCl measured at 25 °C in 100 mM HEPES pH 7.2 using refractometry (a) or SPR (b). All SPR measurements were performed on streptavidin-coated surface.

RIIs were determined from the slopes of the fitted straight lines. Data shown are from two independent sets of samples for each analyte.

3.4. Determination of active target ($c_a$)

In order to correctly calculate the theoretical $R_{max}$ value (Eq. 2), the fraction of active target, i.e. the immobilized molecules actually available for binding, $c_a$, must be determined. For that purpose, complementary DNA strands were injected to the immobilized TBA or TBAcont. It was assumed, that due to strong affinity, all available target will bind its complementary strand and its SPR response, $R_{max}$ given by Eq. 2, would thus correlate with the fraction available. Since in this case the two interacting molecules, DNA, are of the same type, their RIIs are considered identical(Davis and Wilson, 2000; Di Primo and Lebars, 2007; Tumolo et al., 2004) and therefore their ratio equal to 1. The stoichiometry of the interaction is also 1 and Eq. 2 can be, therefore, reduced to:

$$R_{max} = R_1 \times \frac{Mw_2}{Mw_1} \qquad (4)$$

The fraction of active target is then obtained from the ratio of measured $R_{max}$ and the theoretical $R_{max}$ value from Eq. 4.

To investigate how target density influences target availability, three different degrees of surface coverage by target, $R_1$, approximately 50, 500 and 1000 RU (exact values shown in Table 1), were produced on individual channels of a streptavidin coated sensor chip. For each $R_1$, theoretical $R_{max}$ was calculated. Then, 50, 500, and 5000 nM of complementary strand was injected consecutively using the SCK method (Karlsson et al., 2006; Palau and Di Primo, 2012). This method was developed to avoid repeated exposure to the harsh regeneration conditions that sometimes are needed to disrupt strong interactions, which could conceivably alter the surface and/or the immobilized target. In addition, this method makes the runs shorter compared to the classical multi-cycle kinetics method. The binding experiments were performed in a buffer with no potassium ions so that both TBA and TBAcont could interact with their complementary strand with no interference for the former with G4 formation. Three experiments were performed on the same sensor chip. The recorded sensorgrams are shown in Fig. S4 (Supplementary material) for TBA (Fig. S4a, c, e) and TBAcont (Fig. S4b, d, f). Fitting of the measured data to a Langmuir 1:1 model of interaction produced experimental $R_{max}$ values which are summarised in Table 1.

The results revealed that a surprisingly low fraction of the immobilized molecules was actually available for binding. Discrepancy between the fraction of active molecules of the two studied ODNs was also observed, generally about ½ for TBA and between 1/3 and ½ for TBAcont. The fraction available was also shown to decrease with increasing surface density of target. For TBA, twenty-fold increase between the lowest and the highest tested target density reduced the fraction of active molecules from 0.586 to 0.471. The observed decline was more pronounced for TBAcont, where the active fraction fell from 0.499 to 0.319 under the same conditions.

Additionally, injections of complementary strands were also performed on the sensor chips that were used for the interaction experiments with $K^+$ binding to TBA (section 3.5). With a target density $R_1$ of ca. 1000 RU used for these experiments, the fractions of active target were equal to $0.529 \pm 0.059$ for TBA and $0.357 \pm 0.056$ for TBAcont, in good agreement with the values obtained from the experiments detailed above on other sensor chips.

**Table 1.** Theoretical and measured $R_{max}$ for given densities, $R_1$, of immobilized TBA and TBAcont interacting with their respective complementary strands.

| ODN | $R_1$ (RU) | $R_{max}$ (RU) theoretical | $R_{max}$ (RU) measured | Ratio[a] |
|---|---|---|---|---|
| TBA | 47 | 31.2 | $18.3 \pm 0.2$ | $0.586 \pm 0.007$ |
|  | 522 | 347.9 | $180.4 \pm 2.2$ | $0.518 \pm 0.006$ |
|  | 1000 | 666.5 | $313.9 \pm 2.9$ | $0.471 \pm 0.004$ |
| TBAcont | 49 | 33.5 | $16.7 \pm 0.8$ | $0.499 \pm 0.025$ |
|  | 506 | 344.4 | $148.9 \pm 2.6$ | $0.432 \pm 0.007$ |
|  | 1001 | 680.8 | $217.5 \pm 3.4$ | $0.319 \pm 0.005$ |

[a]The ratio, $R_{max}$ measured to $R_{max}$ theoretical, gives the fraction of active target $c_a$

3.5. Experimental measurement of $R_{max}$ for $K^+$ binding to TBA

The experimental maximum SPR signal response, $R_{max}$, of $K^+$ ions binding to TBA was determined from injections of KCl solutions to biotin-TEG-TBA and biotin-TEG-TBAcont immobilized on streptavidin coated sensor chip as described in the experimental section.

The $R_{max}$ corresponding to the specific binding of $K^+$ to TBA cannot be obtained directly from the measured data of TBA interacting with $K^+$, because the raw signal carries contributions not only from the specific binding facilitating folding of the G4 but also from the nonspecific loose associations of the cations with the negatively charged sugar-phosphate backbone of the ODN. To resolve this issue, the binding experiments were performed also with TBAcont, where no specific binding was expected, under the assumption that the signal contribution arising from the nonspecific interaction would be the same for the linear TBAcont as for the G4 TBA. Hence, the signal responses recorded for TBAcont were subtracted from those for TBA, producing data purportedly corresponding to the specific binding of $K^+$ by TBA.

Representative sensorgrams for the interaction of the ODNs with $K^+$ are presented in Fig. 3. Sensorgrams of TBA (Fig. 3a) clearly exhibit typical binding response with fast dissociation kinetics, arising from the interaction with $K^+$. On the other hand, sensorgrams of TBAcont display almost square profile with only slight curvature in the association phase (Fig. 3b), too fast to be measured by the instrument. Evidently, despite the lack of specific binding, electrostatic interactions of $K^+$ with the control strand still induces a considerable signal response, roughly half of the total signal observed for TBA. The SPR signal response was evaluated from the sensorgrams at 5 second before the end of injection (Fig. 3c). As expected from a non-specific interaction signal wth the control ODN, TBAcont, exhibited linear response to the increasing KCl concentration in the whole studied range, whereas for TBA, deviations from linearity were observed due to the contribution of specific binding. Since the amount of immobilized TBA and TBAcont, $R_1$, was never exactly the same, the signal of TBAcont was corrected to correspond to the $R_1$ value of TBA prior to its subtraction from the signal of TBA, in order to accurately compare the two responses. This correction, however, caused only minute changes to the resulting data. Subtraction of signal for TBAcont from signal of TBA produced a saturation curve, which was then fitted to Eq. 3 to acquire the $R_{max}$ value (De Crescenzo et al., 2008) (Fig. 3c). For the average $R_1$ value of $1033 \pm 50$ from all performed experiments, the $R_{max}$ determined this way was $30.9 \pm 3.5$. In addition, $K_D$ was also determined from the fit as $3.4 \pm 0.9$ mM. This value agrees well with the $K_D$ acquired from CD titration, $1.6 \pm 0.3$ mM. Thus, it was shown that immobilization did not significantly alter the affinity of TBA towards $K^+$ ions.

Alternatively, full sensorgrams of TBAcont were subtracted from those of TBA. The resulting set of curves was then directly fitted to a 1:1 Langmuir model using BIAevaluation software (Biacore) to assess the $R_{max}$ and the $K_D$ (Fig. 3d). The quality of the fits was affected by spikes present in the subtracted sensorgrams. These are simply artefacts due to the serial arrangement of the flow channels carrying TBAcont and TBA, causing slight time shift of the beginning of injection between the two channels. Nonetheless, this approach gave an $R_{max}$ and $K_D$ equal to $30.3 \pm 3.5$ and $3.4 \pm 0.9$ mM, respectively, in excellent agreement with the values obtained by steady-state analysis ($30.9 \pm 3.5$ and $3.4 \pm 0.9$ mM).

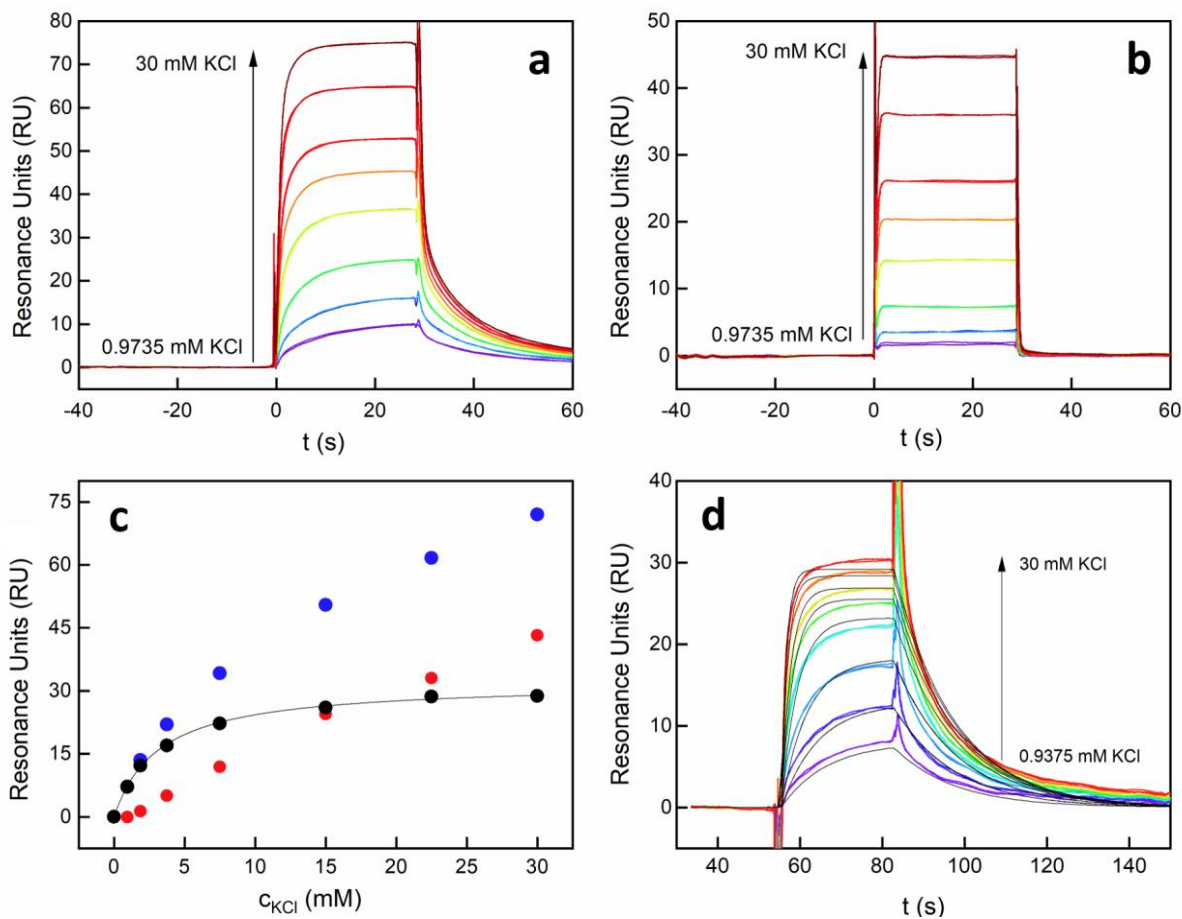

**Figure 3.** SPR analysis of potassium ions binding to TBA. KCl samples prepared in 100 mM HEPES buffer, pH 7.2, were injected at increasing concentrations over approximately 1000 RU of TBA (a) or TBAcont (b). Dependence of the SPR signal of the interaction of KCl (c) with TBA (blue) and TBAcont (red) at 25 s after injection and fitted saturation curve obtained by subtraction of TBAcont data from TBA ones (black). Sensorgrams obtained by subtraction of sensorgrams of TBAcont from those of TBA (d) fitted to a 1:1 Langmuir model of interaction in BIAevaluation Software (Biacore).

3.6. Comparison of experimental and theoretical $R_{max}$

In a series of experiments described above, individual terms of Eq. 2 were determined and are summarised in Table 2. Refractive index increment of TBA ($RII_1$ in Eq. 2) could not be determined unequivocally and, therefore, all relevant applicable values are listed. From the measured values, $R_{max}$ was calculated according to Eq. 2. The calculation was performed for all values of $RII_1$ of TBA from Table 2. Similarly, since the relevance of the measured value of active target available, $c_a$ is questionable (discussed below), the calculation was performed also without including this term. Results of these calculations are summarised in Table 3. The calculated $R_{max}$ values were compared to the experimental $R_{max}$ determined by SPR (section 3.5. also listed in Table 3). Depending on whether different availability of TBA and the subtracted TBAcont is weighted in or not, the experimental $R_{max}$ was determined as either 16.7 or 30.9, respectively (Table 3), without affecting the $K_D$ for potassium binding (Supplementary material, Fig. S5). On the other hand, the calculated $R_{max}$ values range from 1.61 to 5.65 depending on the applied value of $RII_1$ and $c_a$. Thus, it is evident from Table 3 that in all instances a clear discrepancy exists between the experimentally measured and calculated $R_{max}$ values.

**Table 2.** Measured values of individual terms of Eq. 2.

| $R_1{}^a$ (RU) | $RII_1$ (ml g-1) | | | | $RII_2$ (ml g$^{-1}$) | $c_a{}^b$ |
|---|---|---|---|---|---|---|
| | SPR | | Refractometry | | | |
| | KCl 30 mM | no KCl | KCl 30 mM | no KCl | | |
| 1033 | 0.164 | 0.210 | 0.222 | 0.268 | 0.120 | 0.471 |

$^a$Indexes 1 and 2 refer to TBA (5295 g mol$^{-1}$) and K$^+$ (39 g mol$^{-1}$), respectively. $^b$Fraction of the active target. The stoichiometry S (not shown) of K$^+$ binding to TBA is 1

**Table 3.** $R_{max}$ calculated from Eq. 2 for all various determined RII values of TBA ($RII_1$) and experimental $R_{max}$.

| $RII_1$ (ml g$^{-1}$) | Calculated $R_{max}$ (RU) | |
|---|---|---|
| | $c_a$ not considered | $c_a$ considered |
| 0.164 | 5.56 | 2.62 |
| 0.210 | 4.36 | 2.05 |
| 0.222 | 4.12 | 1.94 |
| 0.268 | 3.41 | 1.61 |
| | Experimental $R_{max}$ (RU) | |
| | raw data | corrected data |
| | 30.9 | 16.7 |

## 4. Discussion

### 4.1. Conformational change model

The molecule selected as a conformational change model was the ODN TBA which in the presence of K$^+$ ions is able to fold into a antiparallel G4 structure (Macaya et al., 1993; Wang et al., 1993). A G4 nucleic acid was chosen for this purpose due to its relative simplicity and well-defined K$^+$-driven conformational transition, compared to proteins which were used in most publications claiming to detect structural changes using SPR (Boussaad et al., 2000; Flatmark et al., 2001; Geitmann and Danielson, 2004; Hall et al., 2011; Hsieh et al., 2002; Mannen et al., 2001; Salamon et al., 1994; Sota et al., 1998; Zako et al., 2001). Consequently, unwanted effects often encountered with proteins can be avoided, such as aggregation, target heterogeneity, partial denaturation, or secondary binding sites which all can affect the final SPR signal. Specifically, TBA was chosen as one of the best described G4s that also consists of only two G-tetrads binding a single potassium ion between them, so that a simple 1:1 interaction model (Eq. 3) can be used in data analysis. One drawback of using this G4 is its low molecular weight and, more importantly, the low molecular weight of the analyte ($K^+$), generally expected to produce low SPR response. Low signal can be compensated for by using higher target surface coverage ($R_1$) but only if instruments are sensitive enough to detect low molecular weight compounds. According to the manufacturer, Biacore T200 instruments have no molecular weight detection limit. However, for kinetic analyses high density have been always discouraged as it may negatively affect the interaction (Schasfoort, 2017).

For SPR experiments, TBA biotinylated through TEG linker was used. CD spectroscopy revealed that in conditions used in SPR experiments this covalent conjugate of TBA folds into G4 with increasing concentration of KCl, as expected. Contrarily, CD spectra of G4 non-forming TBAcont displayed no response to added KCl, although presence of clear bands in the spectra indicate non-random conformation of this ODN. Even though the biotin-TEG tail lowered the affinity of TBA to $K^+$ by about one order of magnitude, the used concentration range of KCl was still suitable to study the interaction. Similarly, the covalent modification was shown to destabilize TBA with a $T_m$ value decreasing of about 7 °C. Nonetheless, its thermal stability was still sufficient for the purpose of this study.

4.2. Evaluation of $R_{max}$

With the experimental model validated, the terms of Eq. 2 were collected. Some terms, namely $R_1$, the amount of TBA immobilized, $MW_1$, $MW_2$ and the stoichiometry of the reaction S, did not require experimental assertion. The amount of immobilised target $R_1$ is a parameter that can be controlled by regulating concentration and/or injection time of the biotinylated species over streptavidin-functionalised sensor chip. Since the experiments with varying amount of immobilized target (section 3.4.) revealed only slight negative impact of higher target density on its availability for binding, the highest tested value (≈1000 RU) was selected for the interaction experiment to achieve the highest response. The average $R_1$ from all experiments used for calculating $R_{max}$ was equal to 1033 RU.

The term $MW_2$ corresponds to molecular weight of the analyte, i.e. potassium ion (39 g mol$^{-1}$). For the term $MW_1$, corresponding to the immobilized target, molecular weight of the whole covalent conjugate biotin-TEG-TBA (5295 g mol$^{-1}$) was considered.

Stoichiometry of the interaction between $K^+$ and TBA was considered to be 1:1 (S = 1 in Eq. 2) based on the single specifically bound ion in the central cavity of the G4. It is important to note, that this model is simplified as it assumes that nonspecific electrostatic interaction of $K^+$ with a sugar-phosphate backbone of a G4 would be the same as with a linear ODN of the same length and therefore can be eliminated by subtraction of signal of such control. However, the average number of $K^+$ ions loosely associated with a compact globular G4 structure and an unfolded single-strand at any given time is likely not the same, due to different surface charge density/distribution of either form (Gou et al., 2010; Huang et al., 2007; Mimura et al., 2021). Moreover, in the sequence of TBAcont, four guanines of the original TBA sequence were replaced by adenines, possessing smaller dipole moment (Burda et al., 2012; Šponer et al., 1996) which could likewise affect the surrounding ionic environment. Nonetheless, the number of negatively charged phosphates remains the main factor determining the number of cations associated to an ODN. Lastly, based on the results from section 3.6 roughly five more $K^+$ ions would have to be bound to TBA than to TBAcont in order to compensate the difference between experimental and calculated $R_{max}$.

Knowledge of the ratio of RIIs of the interacting molecules is critical for the analysis of SPR signal in systems where the two interaction partners are chemically very different and the ratio is likely to be far from 1 (Davis and Wilson, 2000; Tumolo et al., 2004). In spite of its importance, this parameter is commonly neglected in SPR studies claiming to detect conformational changes of biomolecules upon interaction with small analytes. RIIs were determined experimentally using refractometry as well as SPR. Each of the used methods utilised a different wavelength for refractive index measurement. However, this is not expected to influence the obtained values significantly. Moreover, in Eq. 2, ratio of the RIIs is used, effectively cancelling any effects resulting from the choice of a method. Although the actual analyte in SPR experiments is the $K^+$ ion, $RII_2$ was determined for KCl, since it is impossible to prepare a solution of the isolated cation. Nonetheless, it should be noted that this approximation might slightly influence the final result.

Unlike for $RII_2$, where the values obtained by either method were similar and their average was used, assorted values were acquired for $RII_1$ of TBA. Firstly, the presence of KCl was shown to cause a non-negligible change of the $RII_1$ value. It is, however, unclear whether this change stems from the KCl-induced folding of TBA or from the different ionic environment. Regardless of the reason, if the RII of TBA is dependent on KCl concentration and thus changes in the course of the SPR interaction experiment, selecting a suitable $RII_1$ value to plug into Eq. 2 becomes problematic. Therefore, the $R_{max}$ calculation had to be done for both limit values, 0 and 30 mM KCl. Furthermore, measurement of $RII_1$ using SPR on streptavidin-coated sensor chip produced different values from refractometric measurements. It is very likely that these differences emerge from interaction of the injected TBA with the complex immobilised layer on the surface of the sensor chip (carboxymethyldextran, streptavidin), especially in the presence of 30 mM KCl as can be seen from the corresponding curves (Supplementary material, Fig. S3d). It is feasible that the value of RII of TBA measured on plain gold surface would more closely approximate the refractometry values, as was seen with KCl (Supplementary material, Fig. S3a - b) and also with other biomolecules (Supplementary material, Fig.

S6-S7, Table S2). This discrepancy between methods further complicates the issue of selecting the accurate value for $RII_1$. On one hand, direct refractometry of samples in solution should be considered the benchmark method for determination of refractive indexes. On the other hand, streptavidin-coated sensor chips were used for the actual SPR interaction experiments of interest, so it can be argued that the value acquired by SPR on the same surface is more suitable for the present application. Ultimately, the choice of the most accurate value, albeit noteworthy, was not critical for the objective of this study. All obtained values were considered in the $R_{max}$ calculation and, as is evident from Table 3, neither one fundamentally impacts the final outcome of the calculation.

The fraction of immobilised target available for $K^+$ binding ($c_a$) determined by injecting the complementary strands to both TBA and TBAcont was unexpectedly low. The marked difference in this value for the two studied ODNs was also surprising. One possible reason for the observed low availability may be steric hinderance from neighbouring strands caused by high target density ($\approx$ 1000 RU). However, twentyfold decrease in surface density of TBA caused only marginal increase in its availability. For TBAcont, this increase was slightly larger, possibly reflecting bigger contribution of the steric effect to availability of this ODN. Another reason for the low response could be target loss between immobilization and the experiment caused by degradation of the DNA. The possibility of target heterogeneity was excluded by quality control of the sample by gel electrophoresis and also due to uniformly oriented immobilisation of target through streptavidin-biotin coupling. Overall, this result points out the importance of target availability determination not only for the purpose of this study but for the use of the SPR technique in general, in particular and obviously for stoichiometry measurements. Although the $c_a$ parameter is needed to calculate the $R_{max}$ value, we were operating under the assumption that the availability of TBA for $K^+$ was the same as for its complementary strand. However, this assumption may likely be untrue due to the vastly different chemical nature of the two species. It is feasible that the much smaller cation would suffer from less steric interference and thus be able to access more of the immobilised TBA strands. Due to this debatable relevancy of the measured $c_a$ value, $R_{max}$ calculations were performed also without taking it into account.

About one half of the raw SPR signal of $K^+$ binding to TBA was attributed to nonspecific electrostatic interactions. The remaining signal was used to determine experimental value of $R_{max}$. Comparison of this value and the $R_{max}$ calculated from the formerly discussed parameters revealed a significant discrepancy which persisted regardless of the $RII_1$ value applied or whether target availability was considered or not. Even though the determined $c_a$ might be unrealistically low as discussed above, even consideration of full availability still produced calculated $R_{max}$ 3 to 9 times lower than experimental. Substantial difference in predicted and measured signal response upon binding of a single ion was previously detected and attributed to conformational change of the immobilised target protein by Gestwicki et al. (Gestwicki et al., 2001), who however, completely disregarded the difference of RIIs of the two species in comparison of predicted and detected signals. In our work, all known contributing parameters to SPR signal were considered, which still could compensate maximally for about 18 % of the measured signal. Simultaneously, conformational change of TBA was confirmed by CD to occur under the used conditions. This allows us to attribute the uncompensated SPR signal to refractive index change as a result of conformational change of TBA occurring on the sensor chip surface.

## 5. Conclusions

This work set out to deliberately answer the question whether conformational change of a biomolecule immobilised on a sensor chip surface during SPR experiments contributes to the recorded SPR signal. For this purpose, all known contributing parameters to the signal were thoroughly experimentally inspected to assess the theoretical maximum response $R_{max}$ of the model G4-forming oligonucleotide TBA binding $K^+$ ion. Various values of refractive index increment of TBA were found in the presence or absence of KCl or depending on the method used for its determination. All collected values were, therefore, considered in $R_{max}$ calculation. Additionally, strikingly low fraction of immobilised TBA on an SPR sensor chip available for interaction with its binding partner was determined, which highlighted the impact this parameter could have on the final SPR signal. Experimental measurement of $R_{max}$ revealed significant contribution of nonspecific interactions between TBA and $K^+$ to the detected SPR response. Experimentally determined $R_{max}$ was shown to be several times higher than the values predicted by calculation after consideration of all known contributing factors, as a result of TBA folding into G4 in the course of the experiment. Therefore, we conclude that conformational changes can contribute to the SPR signal measured on Biacore instruments, the most widely used biosensor to characterize molecular interactions.

**Acknowledgments**

CD is very grateful to Dr. Jean-Louis Mergny for suggesting DD to come to the IECB and to Dr. Miroslav Fojta from the Institute of Biophysics of the Czech Academy of Sciences for allowing DD to stay in Pessac. This work has been supported by the SYMBIT project reg. no. CZ.02.1.01/0.0/0.0/15_003/0000477 financed from the ERDF and by the INSERM U1212 - CNRS UMR 3033 ARNA laboratory. We gratefully thank Dr. Mikayel Aznauryan for careful reading of the manuscript. We thank Alysson Duchalet for performing preliminary experiments and Lætitia Minder from the Structural Biophysico-Chemistry plateform of the European Institute of Chemistry and Biology (IECB, UAR 3033 - US001) for technical assistance with the Biacore T200. The instrument was acquired in 2011 with the support of the Conseil Régional d'Aquitaine, the GIS-IBiSA and the Cellule Hôtels à Projets of the CNRS.